%% file: paper_v.13.tex
\documentclass[aps,prl,twocolumn,showpacs,groupedaddress]{revtex4}  
\usepackage{graphicx}
\usepackage{dcolumn}
\usepackage{bm}
\usepackage{amssymb}   

\newcommand{\met}       {\mbox{$\not\!\!E_T$}}

\begin{document}
\hspace{5.2in} \mbox{FERMILAB-PUB-05-087-E}
\title{Measurement of the \mbox{\boldmath$t\bar{t}$}  Production Cross Section 
in \mbox{\boldmath$p\bar{p}$} Collisions at \mbox{\boldmath$\sqrt{s}=1.96$}~TeV using 
Lepton + Jets Events with Lifetime \mbox{\boldmath$b$}-tagging}

\input list_of_authors_r2.tex
\date{April 28, 2005} 
           
\begin{abstract}
We present a measurement of the top quark pair ($t\bar{t}$) production cross 
section ($\sigma_{t\bar{t}}$) in $p\bar{p}$ collisions at $\sqrt{s}=1.96$~TeV using 
230 pb$^{-1}$ of data collected by the D\O\  experiment at the Fermilab Tevatron Collider. 
We select events with one charged lepton (electron or muon), missing
transverse energy, and jets in the final state. We employ lifetime-based
$b$-jet identification techniques to further enhance the $t\bar{t}$ purity of
the selected sample. For a top quark mass of 175 GeV, we measure 
$\sigma_{t\bar{t}}=8.6^{+1.6}_{-1.5}\;({\rm stat.+syst.})\pm 0.6\;({\rm lumi.})$ pb, 
in agreement with the standard model expectation.
\end{abstract}

\maketitle

The top quark was discovered at the Fermilab Tevatron Collider 
by the CDF and D\O\ collaborations \cite{topdisc} in $p\bar{p}$ 
collisions at $\sqrt{s} = 1.8$ TeV based on about 50 pb$^{-1}$ of data per
experiment. The increased statistics and higher collision energy of 
$\sqrt{s} = 1.96$ TeV of Tevatron Run II allow more precise measurements of top quark
properties, including its production and decay characteristics. Theoretical
calculations performed within the framework of the standard model (SM) predict
the $t\bar{t}$ production cross section ($\sigma_{t\bar{t}}$) with an uncertainty 
of less than 15\%~\cite{theorxsect}. A significant deviation from this prediction would 
signal the presence of physics beyond the SM, such as $t\bar{t}$ resonant production~\cite{theorNP}.
The CDF and D\O\ collaborations have previously reported measurements of $\sigma_{t\bar{t}}$
at $\sqrt{s}=1.8$~TeV~\cite{TeVRunI}. Recent measurements at $\sqrt{s}=1.96$~TeV 
by the CDF~\cite{CDF} and D\O\ \cite{ljetstopo_paper} collaborations agree with the 
SM prediction within their experimental uncertainties.
  
In the SM, the top quark decays to a $W$ boson and $b$ quark with 
a branching ratio of $\approx 100$\%. The lepton+jets final state results
from the leptonic decay of one of the $W$ bosons and the hadronic decay
of the other. The event signature is one lepton
with high transverse momentum, large transverse energy imbalance (\met)
due to the undetected neutrino, and four jets, two of which result from  
hadronization of the $b$ quarks.

In this Letter, we report the measurement of $\sigma_{t\bar{t}}$
in the lepton (electron or muon) plus jets channel using $b$-jet 
identification ($b$-tagging) techniques exploiting the long lifetime of $B$ hadrons.
The data were collected by the D\O\ experiment from August 2002 through March 2004,  
and correspond to an integrated luminosity of $226\pm15$~pb${}^{-1}$ 
($229\pm15$~pb${}^{-1}$) in the electron (muon) sample.

The D\O\ detector includes a tracking system, calorimeters, and a muon
spectrometer~\cite{d0det}. The tracking system consists of a silicon microstrip
tracker (SMT) and a central fiber tracker (CFT), both located inside a 2~T
superconducting solenoid. The tracker design provides  efficient charged
particle measurements in the pseudorapidity region $|\eta| < 3$~\cite{eta}. The
SMT strip  pitch of 50--80~$\mu$m allows a precise reconstruction of the
primary interaction vertex (PV) and an accurate determination of the impact
parameter of a track relative to the PV~\cite{ip}, which are the key
components of the lifetime-based $b$-jet tagging algorithms. The PV
is required to be within the SMT fiducial volume and consist of at least 3 tracks. The calorimeter
consists of a central section (CC) covering $|\eta|<1.1$, and two end
calorimeters (EC) extending the coverage to  $|\eta|\approx4.2$.  The muon system
surrounds the calorimeter and consists of three layers of tracking detectors 
and two layers of scintillators~\cite{muon_detector}. A 1.8 T iron
toroidal magnet is located outside the innermost layer of the muon detector.
The luminosity is calculated from the rate for
{\mbox{$p\bar p$}}\ inelastic collisions detected using two hodoscopes of scintillation
counters mounted close to the beam pipe on the front surfaces of the EC
calorimeters.

We select data samples in the electron and muon channels
by requiring an isolated electron with $p_T>20$~GeV and
$|\eta|<1.1$, or an isolated muon with $p_T>20$~GeV and $|\eta|<2.0$. More
details on the lepton identification as well as trigger requirements are reported
elsewhere~\cite{ljetstopo_paper}. In both channels, we require {\met}
to exceed $20$~GeV and not be collinear with the lepton direction in the
transverse plane. These $W$ boson candidate events must be accompanied by one 
or more jets with $p_T>15$~GeV and rapidity $|y|<2.5$~\cite{eta}. Jets are 
defined using a cone algorithm with radius  $\Delta{\cal R}=0.5$~\cite{jet}.  
We classify the selected events according to their jet multiplicity. 
Events with 3 or $\geq 4$ jets are expected to
be enriched in $t\bar{t}$ signal, whereas events with only 1 or 2 jets
are expected to be dominated by background. We use the former to estimate
$\sigma_{t\bar{t}}$, and the latter to verify the background normalization procedure.

The main background in this analysis is the production of $W$
bosons in association with jets ($W$+jets), with the $W$ boson decaying 
leptonically. In most cases, the jets accompanying the $W$ boson
originate from light ($u$, $d$, $s$) quarks and gluons ($W$+light jets). Depending on the jet 
multiplicity, between $2$\% and $14$\% of $W$+jets events contain heavy flavor jets resulting 
from gluon splitting into $b\bar{b}$ or $c\bar{c}$ ($Wb\bar{b}$ or $Wc\bar{c}$, 
respectively), while in about $5$\% of events, a single $c$ quark is present in
the final state as a result of the $W$ boson radiated from an $s$ quark from
the proton's or antiproton's sea ($Wc$). 
A sizeable background arises from strong production of two or more jets 
(``multijets''), 
with one of the jets misidentified as a lepton and accompanied by large {\met} 
resulting from mismeasurements of jet energies.
Significantly smaller contributions to the background arise from
single top, $Z$+jets, and weak diboson ($WW$, $WZ$ and $ZZ$) production. 
Only a small fraction of the background events contain $b$ or
$c$-quark jets in the final state. As a consequence, the signal-to-background ratio 
is significantly enhanced when at least one jet is identified as a $b$-quark jet.

We use a secondary vertex tagging (SVT) algorithm to identify $b$-quark jets. 
Secondary vertices are reconstructed from two or more tracks satisfying the following
requirements: $p_T>1$ GeV, $\geq 1$ hits in the SMT layers and impact parameter significance
$d_{ca}/\sigma_{d_{ca}}>3.5$~\cite{ip}. Tracks identified as arising from $K^0_S$ or $\Lambda$ 
decays or from $\gamma$ conversions are not considered. If the secondary vertex reconstructed 
within a jet has a decay length significance $L_{xy}/\sigma_{L_{xy}}>7$~\cite{dl}, 
the jet is tagged as a $b$-quark jet.  Events with
exactly 1 ($\geq 2$) tagged jets are referred to as single-tag (double-tag)
events. We treat single-tag and double-tag events separately because of their
different signal-to-background ratios. 

Secondary vertices with $L_{xy}/\sigma_{L_{xy}}<-7$ appear due to finite 
resolution of their characteristics after reconstruction, and define the ``negative tagging
rate''. The negative tagging rate is used to estimate the  probability for
misidentifying a  light flavor ($u$, $d$, $s$ quark or gluon) jet as a
$b$-quark jet (the ``mis-tagging rate'').

\begin{figure}[t]
\resizebox{7cm}{7cm}{\includegraphics{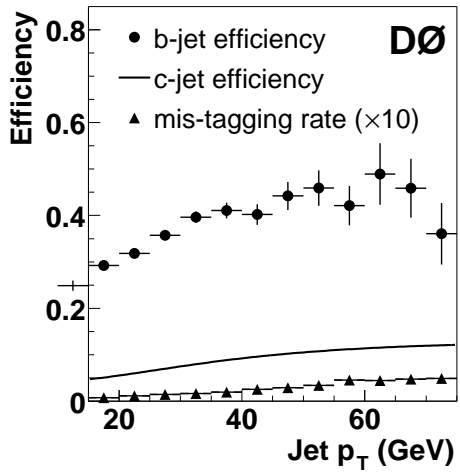}}
\caption{\label{fig:tag_eff} Measured $b$-tagging efficiency (circles) and
mis-tagging rate (triangles), and estimated $c$-tagging efficiency (solid line) 
as a function of jet $p_T$. } \end{figure}

We estimate both the $b$-tagging efficiency and the mis-tagging rate using
jets with $\geq 2$ tracks satisfying less stringent requirements than those
for SVT. In particular, the $p_T$ cut is reduced from 1 GeV to 0.5 GeV for
all but the highest $p_T$ track, and no cut on $d_{ca}/\sigma_{d_{ca}}$ 
of the tracks is made. These requirements have an efficiency per jet 
$> 80$\% for $p_T > 30$ GeV and integrated over $y$.
We measure the $b$-tagging efficiency in a data sample of dijet events 
with enhanced heavy flavor content by requiring a jet
with an associated muon at high transverse momentum relative to the jet axis. 
By comparing the SVT and muon-tagged jet samples, the tagging efficiency
for semileptonic $b$-quark decays (``semileptonic $b$-tagging efficiency'') can 
be inferred. We make use of a Monte Carlo (MC) simulation to
further correct the measured efficiency to the tagging efficiency for inclusive
$b$-quark decays. We estimate the $c$-tagging efficiency from the same simulation,
corrected by a scale factor defined as the ratio of the semileptonic $b$-tagging 
efficiency measured in data to that measured in 
the simulation. We estimate the mis-tagging rate from the negative tagging rate
measured in dijet events, corrected for the contribution of heavy-flavor jets and
the presence of long-lived particles in light-flavor jets. 
Figure~\ref{fig:tag_eff} shows the $b$-tagging efficiency, $c$-tagging efficiency
and mis-tagging rate as a function of jet $p_T$.

We simulate $t\bar{t}$ production, and all background
processes except multijets, using {\sc alpgen}~\cite{ALPGEN} to generate the
parton-level processes, and {\sc pythia}~\cite{PYTHIA} to provide fragmentation and 
to decay all unstable
particles except $B$ hadrons and $\tau$ leptons, which are modeled via {\sc
evtgen}~\cite{EVTGEN} and {\sc tauola}~\cite{TAUOLA}, respectively. We process the
generated events through the full {\sc geant}-based~\cite{GEANT} D\O\
detector simulation and the same reconstruction program used to process the 
data. We apply small additional smearing to the reconstructed objects to
improve the agreement between the data and the simulation, and
account for remaining discrepancies using 
correction factors derived by comparing the efficiencies measured in 
{\mbox{$ Z\rightarrow \ell^+ \ell^-$}} data to those obtained from the simulation. 
For all processes except the multijets background, we make use of the MC
simulation to compute the total acceptance, applying the trigger, reconstruction
and tagging efficiencies measured using data. The tagging probability for a particular
process depends on the flavor composition of the jets in the final state as
well as on the overall event kinematics. We estimate it by applying the tagging rates 
measured in data to each jet in the simulation, taking into consideration
its flavor, $p_T$, and $y$. In the case of $W$+jets events, we also use the
simulation to estimate the fraction of the different $W$+heavy flavor
subprocesses.

We compute the $t\bar{t}$ acceptance for events with a true electron or muon
arising from a $W\rightarrow \ell \nu$ ($\ell=e,\mu,\tau$) decay,
corresponding to total branching fractions of $17.106$\% and
$17.036$\%~\cite{PDG2002}, respectively, in the electron and muon channels. In
the electron channel, the total acceptance before tagging is estimated to be
$(10.8 \pm 0.8)$\% and $(14.2 \pm 1.7)$\%, for events with
3 and those with $\geq 4$ jets, respectively. The corresponding numbers for the muon
channel are $(9.9 \pm 1.0)$\% and $(14.1 \pm 1.9)$\%. The estimated single-tag
efficiencies are $(43.4 \pm 1.2)$\% and $(45.3 \pm 1.0)$\%
for events with 3 and those with $\geq 4$ jets, respectively. The corresponding 
double-tag efficiencies are $(10.4 \pm 1.0)$\% and $(14.2 \pm 1.3)$\%.

We estimate the number of multijet events from the data for each jet
multiplicity using the matrix method described in Ref.~\cite{ljetstopo_paper},
separately for the samples before and after tagging. Smaller contributions from
single top, $Z$+jets, and diboson production (collectively referred to as ``other bkg'')
are estimated from the simulation, normalized to the next-to-leading
order theoretical cross sections~\cite{NLO_singletop,NLO_diboson}. We also include
under ``other bkg'' the contribution from $t\bar{t}$ with both $W$ bosons
decaying leptonically, assuming the same $\sigma_{t\bar{t}}$ as for $t\bar{t}\rightarrow\ell$+jets. 
We determine the number of tagged $W$+jets events as the product of the number of $W$+jets 
events in data before tagging and the average tagging probability for $W$+jets
events (e.g. $\approx 4$\% for single-tag and $\approx 0.4$\% for double-tag events with
$\geq 4$ jets). The number of $W$+jets events before tagging  is computed as a difference 
between the number of selected  events and the estimated contribution from the rest of
processes (multijets, $t\bar{t}$, and ``other bkg'').

\input{svt_table_PRL.tex}

\begin{figure*}   
\begin{center}
\includegraphics[width=0.4\textwidth]{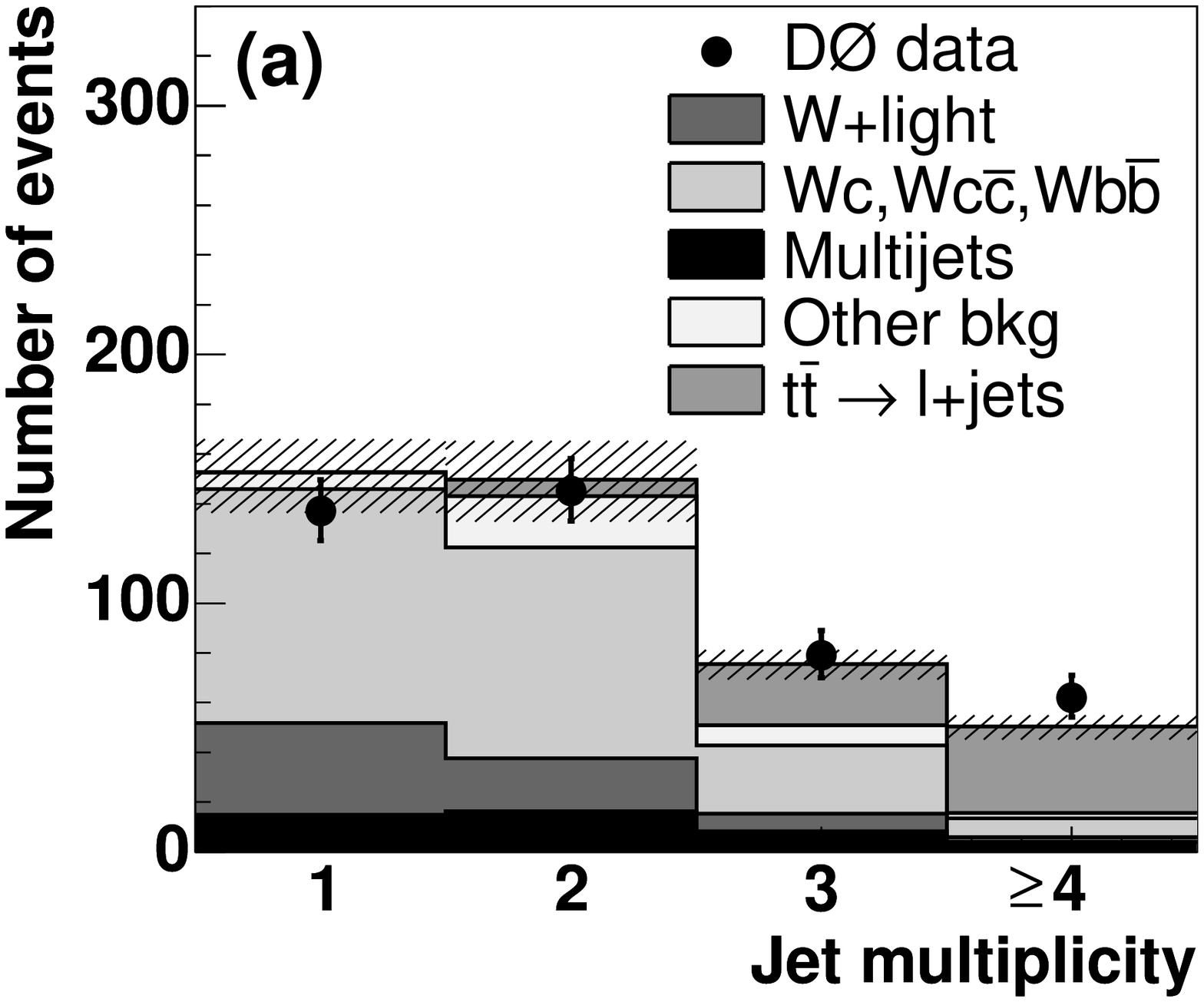}
\includegraphics[width=0.4\textwidth]{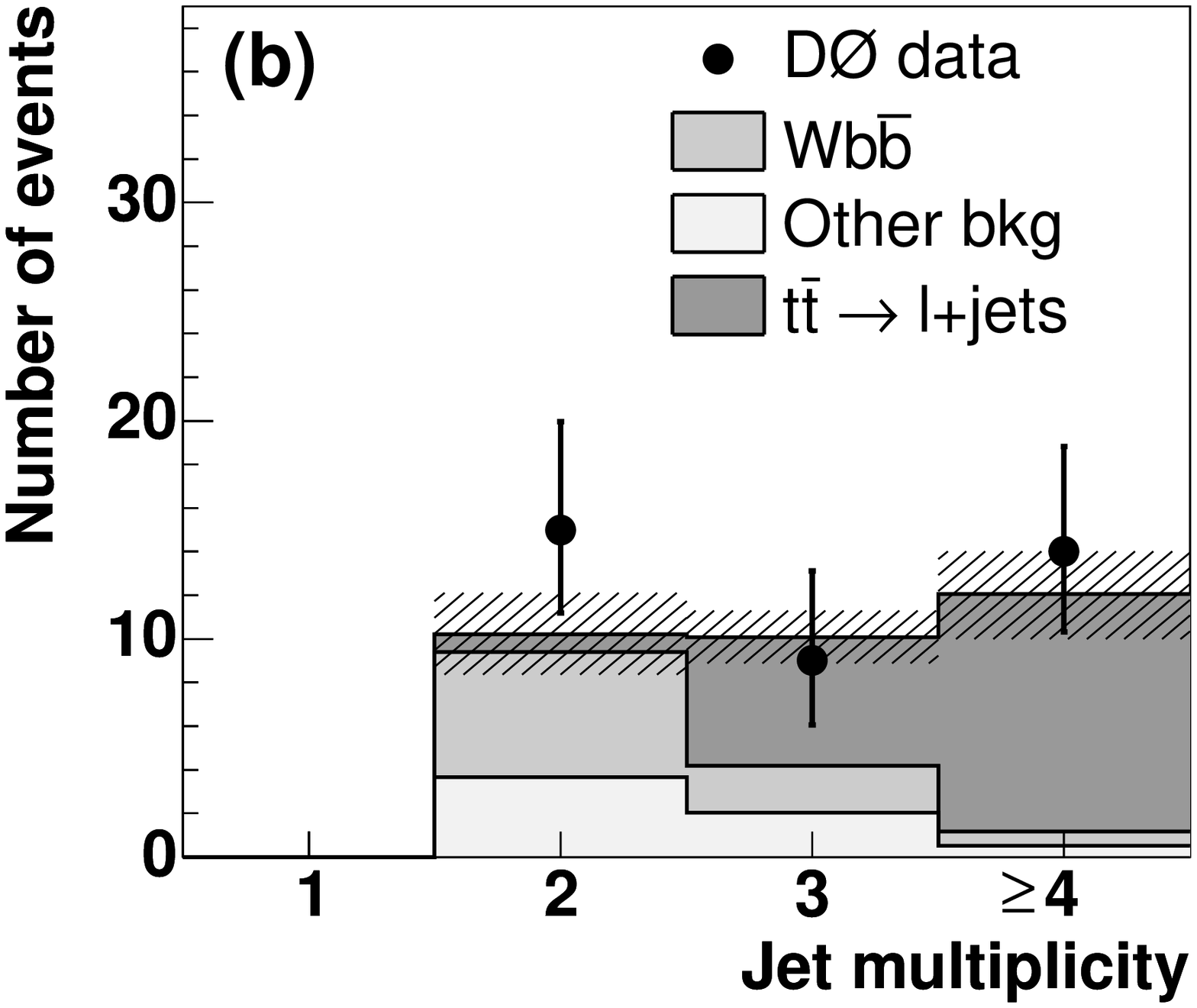}
\end{center}
\caption{\label{fig:SVT} Expected and observed number of (a) single-tag
 and (b) double-tag events. The hatched area represents the total
uncertainty in the expectation.}
\end{figure*}

Tables ~\ref{tab:svt_summary_table_ljets_1}
and~\ref{tab:svt_summary_table_ljets_2} summarize the sample composition for
single-tag and double-tag events, respectively, assuming $\sigma_{t\bar{t}}=7.0$ pb.
Figure~\ref{fig:SVT} shows the observed and expected number of events for each
jet multiplicity. We interpret the excess over the background expectation in 
the third and fourth jet multiplicity bins as the $t\bar{t}$ signal. The good
agreement between observation and expectation in the first and second jet
multiplicity bins validates the background estimation procedure. 

We calculate $\sigma_{t\bar{t}}$ by
maximizing a likelihood function including a Poisson term
for each of the eight independent channels considered:
3 and $\geq 4$ jets, for single- and double-tag
events in the electron and muon channels. At each
step in the maximization, the multijet background in these eight
tagged samples, and the corresponding samples before tagging,
is constrained within errors to the amount determined by the matrix method.
In addition, we include a Gaussian term for each
of the systematic uncertainties considered, following the procedure
described in Ref.~\cite{nuisance_par}. In this approach,
each source of systematic uncertainty is allowed to affect the central
value of the cross section during the maximization procedure, thus
yielding a combined statistical and systematic uncertainty on $\sigma_{t\bar{t}}$.
Assuming a top quark mass ($m_t$) of 175 GeV, we measure 
\begin{eqnarray}
\sigma_{t\bar{t}} = 8.6 ^{+1.6}_{-1.5}\;({\rm stat.+ syst.})\pm
0.6\;({\rm lumi.})\:{\rm pb}, \nonumber
\end{eqnarray}
\noindent in good agreement with the SM prediction of $6.77\pm 0.42$ pb~\cite{theorxsect}.

The contribution due to each individual source
of systematic uncertainty can be estimated by redoing the fit after fixing
all but the corresponding Gaussian term and unfolding the
statistical uncertainty from the resulting total uncertainty. The statistical
uncertainty of $^{+1.2}_{-1.1}$~pb is obtained from the fit where all Gaussian terms are fixed.
\input{syst_table.tex}
As shown in Table~\ref{tab:x_emu}, $b$-jet tagging efficiency, jet energy calibration, 
and background modeling are the leading sources of systematic uncertainty. In addition,
a systematic uncertainty of $6.5$\% from the luminosity measurement~\cite{lumi} has been assigned.
In the top quark mass range of 160 GeV to 190 GeV, the measured cross section 
decreases (increases) by 0.06 pb per 1 GeV shift of $m_t$ above (below) 175 GeV.

We used an alternative $b$-tagging algorithm to cross check this result.
This algorithm relies on counting tracks with significant impact parameter (CSIP) 
with respect to the PV: a jet is tagged if $\geq 2$ ($\geq 3$) associated
tracks have $d_{ca}/\sigma_{d_{ca}}>3$ ($d_{ca}/\sigma_{d_{ca}}>2$).
As compared to SVT, this algorithm has a slightly higher $b$-tagging 
efficiency and about a factor of two higher mis-tagging rate. The measured 
cross section using the CSIP algorithm is
$\sigma_{t\bar{t}} = 7.6^{+1.7}_{-1.4}\;({\rm stat.+ syst.})\pm 0.5\;({\rm lumi.})\:{\rm{pb}}$,
consistent with the SVT result once the existing overlap between both samples 
is taken into account. While we are currently not combining these two results,
the fact that different $b$-tagging techniques are only partially correlated
will be exploited in future analyses to further increase the precision of this
measurement.

In summary, we have measured the $t\bar{t}$ production cross section  
in $p\bar{p}$ interactions at $\sqrt{s} = 1.96$ TeV in the lepton+jets channel
using lifetime $b$-tagging.
Our measurement yields $\sigma_{t\bar{t}} = 8.6^{+1.6}_{-1.5}\;({\rm stat.+ syst.})\pm 0.6\;({\rm lumi.})\:{\rm{pb}}$,
in a good agreement with the SM prediction. 
  
\input acknowledgement_paragraph_r2.tex
\end{document}

%% file: list_of_authors_r2.tex
%
\author{                                                                      
V.M.~Abazov,$^{35}$                                                           
B.~Abbott,$^{72}$                                                             
M.~Abolins,$^{63}$                                                            
B.S.~Acharya,$^{29}$                                                          
M.~Adams,$^{50}$                                                              
T.~Adams,$^{48}$                                                              
M.~Agelou,$^{18}$                                                             
J.-L.~Agram,$^{19}$                                                           
S.H.~Ahn,$^{31}$                                                              
M.~Ahsan,$^{57}$
% ...
\\                                                                            
\vskip 0.30cm                                                                 
\centerline{(D\O\ Collaboration)}                                             
\vskip 0.30cm                                                                 
}                                                                             
\affiliation{                                                                 
\centerline{$^{1}$Universidad de Buenos Aires, Buenos Aires, Argentina}       
\centerline{$^{2}$LAFEX, Centro Brasileiro de Pesquisas F{\'\i}sicas,         
                  Rio de Janeiro, Brazil}                                     
\centerline{$^{3}$Universidade do Estado do Rio de Janeiro,                   
                  Rio de Janeiro, Brazil}                                     
\centerline{$^{4}$Instituto de F\'{\i}sica Te\'orica, Universidade            
                  Estadual Paulista, S\~ao Paulo, Brazil}                     
\centerline{$^{5}$University of Alberta, Edmonton, Alberta, Canada,           
               Simon Fraser University, Burnaby, British Columbia, Canada,}   
\centerline{York University, Toronto, Ontario, Canada, and                    
         McGill University, Montreal, Quebec, Canada}                         
\centerline{$^{6}$Institute of High Energy Physics, Beijing,                  
                  People's Republic of China}                                 
\centerline{$^{7}$University of Science and Technology of China, Hefei,       
                  People's Republic of China}                                 
\centerline{$^{8}$Universidad de los Andes, Bogot\'{a}, Colombia}             
\centerline{$^{9}$Center for Particle Physics, Charles University,            
                  Prague, Czech Republic}                                     
\centerline{$^{10}$Czech Technical University, Prague, Czech Republic}        
\centerline{$^{11}$Institute of Physics, Academy of Sciences, Center          
                  for Particle Physics, Prague, Czech Republic}               
\centerline{$^{12}$Universidad San Francisco de Quito, Quito, Ecuador}        
\centerline{$^{13}$Laboratoire de Physique Corpusculaire, IN2P3-CNRS,         
                 Universit\'e Blaise Pascal, Clermont-Ferrand, France}        
\centerline{$^{14}$Laboratoire de Physique Subatomique et de Cosmologie,      
                  IN2P3-CNRS, Universite de Grenoble 1, Grenoble, France}     
\centerline{$^{15}$CPPM, IN2P3-CNRS, Universit\'e de la M\'editerran\'ee,     
                  Marseille, France}                                          
\centerline{$^{16}$Laboratoire de l'Acc\'el\'erateur Lin\'eaire,              
                  IN2P3-CNRS, Orsay, France}                                  
\centerline{$^{17}$LPNHE, IN2P3-CNRS, Universit\'es Paris VI and VII,         
                  Paris, France}                                              
\centerline{$^{18}$DAPNIA/Service de Physique des Particules, CEA, Saclay,    
                  France}                                                     
\centerline{$^{19}$IReS, IN2P3-CNRS, Universit\'e Louis Pasteur, Strasbourg,  
                France, and Universit\'e de Haute Alsace, Mulhouse, France}   
\centerline{$^{20}$Institut de Physique Nucl\'eaire de Lyon, IN2P3-CNRS,      
                   Universit\'e Claude Bernard, Villeurbanne, France}         
\centerline{$^{21}$III. Physikalisches Institut A, RWTH Aachen,               
                   Aachen, Germany}                                           
\centerline{$^{22}$Physikalisches Institut, Universit{\"a}t Bonn,             
                  Bonn, Germany}                                              
\centerline{$^{23}$Physikalisches Institut, Universit{\"a}t Freiburg,         
                  Freiburg, Germany}                                          
\centerline{$^{24}$Institut f{\"u}r Physik, Universit{\"a}t Mainz,            
                  Mainz, Germany}                                             
\centerline{$^{25}$Ludwig-Maximilians-Universit{\"a}t M{\"u}nchen,            
                   M{\"u}nchen, Germany}                                      
\centerline{$^{26}$Fachbereich Physik, University of Wuppertal,               
                   Wuppertal, Germany}                                        
\centerline{$^{27}$Panjab University, Chandigarh, India}                      
\centerline{$^{28}$Delhi University, Delhi, India}                            
\centerline{$^{29}$Tata Institute of Fundamental Research, Mumbai, India}     
\centerline{$^{30}$University College Dublin, Dublin, Ireland}                
\centerline{$^{31}$Korea Detector Laboratory, Korea University,               
                   Seoul, Korea}                                              
\centerline{$^{32}$CINVESTAV, Mexico City, Mexico}                            
\centerline{$^{33}$FOM-Institute NIKHEF and University of                     
                  Amsterdam/NIKHEF, Amsterdam, The Netherlands}               
\centerline{$^{34}$Radboud University Nijmegen/NIKHEF, Nijmegen, The          
                  Netherlands}                                                
\centerline{$^{35}$Joint Institute for Nuclear Research, Dubna, Russia}       
\centerline{$^{36}$Institute for Theoretical and Experimental Physics,        
                  Moscow, Russia}                                             
\centerline{$^{37}$Moscow State University, Moscow, Russia}                   
\centerline{$^{38}$Institute for High Energy Physics, Protvino, Russia}       
\centerline{$^{39}$Petersburg Nuclear Physics Institute,                      
                   St. Petersburg, Russia}                                    
\centerline{$^{40}$Lund University, Lund, Sweden, Royal Institute of          
                   Technology and Stockholm University, Stockholm,            
                   Sweden, and}                                               
\centerline{Uppsala University, Uppsala, Sweden}                              
\centerline{$^{41}$Lancaster University, Lancaster, United Kingdom}           
\centerline{$^{42}$Imperial College, London, United Kingdom}                  
\centerline{$^{43}$University of Manchester, Manchester, United Kingdom}      
\centerline{$^{44}$University of Arizona, Tucson, Arizona 85721, USA}         
\centerline{$^{45}$Lawrence Berkeley National Laboratory and University of    
                  California, Berkeley, California 94720, USA}                
\centerline{$^{46}$California State University, Fresno, California 93740, USA}
\centerline{$^{47}$University of California, Riverside, California 92521, USA}
\centerline{$^{48}$Florida State University, Tallahassee, Florida 32306, USA} 
\centerline{$^{49}$Fermi National Accelerator Laboratory, Batavia,            
                   Illinois 60510, USA}                                       
\centerline{$^{50}$University of Illinois at Chicago, Chicago,                
                   Illinois 60607, USA}                                       
\centerline{$^{51}$Northern Illinois University, DeKalb, Illinois 60115, USA} 
\centerline{$^{52}$Northwestern University, Evanston, Illinois 60208, USA}    
\centerline{$^{53}$Indiana University, Bloomington, Indiana 47405, USA}       
\centerline{$^{54}$University of Notre Dame, Notre Dame, Indiana 46556, USA}  
\centerline{$^{55}$Iowa State University, Ames, Iowa 50011, USA}              
\centerline{$^{56}$University of Kansas, Lawrence, Kansas 66045, USA}         
\centerline{$^{57}$Kansas State University, Manhattan, Kansas 66506, USA}     
\centerline{$^{58}$Louisiana Tech University, Ruston, Louisiana 71272, USA}   
\centerline{$^{59}$University of Maryland, College Park, Maryland 20742, USA} 
\centerline{$^{60}$Boston University, Boston, Massachusetts 02215, USA}       
\centerline{$^{61}$Northeastern University, Boston, Massachusetts 02115, USA} 
\centerline{$^{62}$University of Michigan, Ann Arbor, Michigan 48109, USA}    
\centerline{$^{63}$Michigan State University, East Lansing, Michigan 48824,   
                   USA}                                                       
\centerline{$^{64}$University of Mississippi, University, Mississippi 38677,  
                   USA}                                                       
\centerline{$^{65}$University of Nebraska, Lincoln, Nebraska 68588, USA}      
\centerline{$^{66}$Princeton University, Princeton, New Jersey 08544, USA}    
\centerline{$^{67}$Columbia University, New York, New York 10027, USA}        
\centerline{$^{68}$University of Rochester, Rochester, New York 14627, USA}   
\centerline{$^{69}$State University of New York, Stony Brook,                 
                   New York 11794, USA}                                       
\centerline{$^{70}$Brookhaven National Laboratory, Upton, New York 11973, USA}
\centerline{$^{71}$Langston University, Langston, Oklahoma 73050, USA}        
\centerline{$^{72}$University of Oklahoma, Norman, Oklahoma 73019, USA}       
\centerline{$^{73}$Brown University, Providence, Rhode Island 02912, USA}     
\centerline{$^{74}$University of Texas, Arlington, Texas 76019, USA}          
\centerline{$^{75}$Southern Methodist University, Dallas, Texas 75275, USA}   
\centerline{$^{76}$Rice University, Houston, Texas 77005, USA}                
\centerline{$^{77}$University of Virginia, Charlottesville, Virginia 22901,   
                   USA}                                                       
\centerline{$^{78}$University of Washington, Seattle, Washington 98195, USA}  
}                                                                             

%% file: svt_table_PRL.tex
\begin{table}[t]
\caption{\label{tab:svt_summary_table_ljets_1}
Summary of observed and expected numbers of events before tagging and
with exactly one jet tagged.}
\begin{tabular*}{\linewidth}{l@{\extracolsep{\fill}}r@{\extracolsep{0pt}}c@{\extracolsep{0pt}}l@{\extracolsep{\fill}}r@{\extracolsep{0pt}}c@{\extracolsep{0pt}}l@{\extracolsep{\fill}}r@{\extracolsep{0pt}}c@{\extracolsep{0pt}}l@{\extracolsep{\fill}}r@{\extracolsep{0pt}}c@{\extracolsep{0pt}}l}
\hline\hline
& $W$+&1&jet & $W$+&2&jets & $W$+&3&jets & $W$+$\ge$&4&jets \\
\hline
&\multicolumn{12}{c}{ Before tagging } \\
\hline
Observed & \multicolumn{3}{c}{14054} & \multicolumn{3}{c}{5502} & \multicolumn{3}{c}{1365} & \multicolumn{3}{c}{367}\\
Multijets & 718&$\pm$&78 & 516&$\pm$&43 & 190&$\pm$&14 & 66&$\pm$&6 \\
\hline
 & \multicolumn{12}{c}{ After tagging } \\
\hline

$W$+light & 36.8&$\pm$&4.0 & 21.4&$\pm$&2.4 & 7.2&$\pm$&0.9 & 1.8&$\pm$&0.3 \\
$Wc$ & 47.8&$\pm$&5.4 & 24.2&$\pm$&2.7 & 5.7&$\pm$&0.7 & 0.8&$\pm$&0.1 \\
$Wc\bar{c}$ & 12.2&$\pm$&3.4 & 17.2&$\pm$&4.8 & 6.6&$\pm$&1.9 & 2.2&$\pm$&0.7 \\
$Wb\bar{b}$ & 33.9&$\pm$&8.7 & 43.2&$\pm$&11.0 & 15.1&$\pm$&3.9 & 4.5&$\pm$&1.3 \\
Multijets & 14.9&$\pm$&1.9 & 16.3&$\pm$&2.1 & 8.3&$\pm$&1.5 & 4.0&$\pm$&1.2 \\
Other bkg & 6.6&$\pm$&0.9 & 20.6&$\pm$&2.1 & 8.2&$\pm$&0.8 & 2.2&$\pm$&0.4 \\
\hline
Total bkg & 152.4&$\pm$&14.8 & 142.9&$\pm$&16.0 & 51.0&$\pm$&5.6 & 15.6&$\pm$&1.7 \\
$t\bar{t}\rightarrow \ell$+jets & 0.4&$\pm$&0.1 & 6.8&$\pm$&1.4 & 24.4&$\pm$&1.7 & 34.8&$\pm$&4.3 \\
\hline
Total expected & 152.8&$\pm$&14.8 & 149.7&$\pm$&16.2 & 75.4&$\pm$&5.9 & 50.4&$\pm$&4.8 \\
Observed & \multicolumn{3}{c}{137} & \multicolumn{3}{c}{145} & \multicolumn{3}{c}{79} & \multicolumn{3}{c}{62}\\
\hline\hline
\end{tabular*}
\end{table}
\begin{table}[t]
\caption{\label{tab:svt_summary_table_ljets_2}
Summary of observed and expected number of events with two or more jets
tagged.}
\begin{tabular*}{\linewidth}{l@{\extracolsep{\fill}}r@{\extracolsep{0pt}}c@{\extracolsep{0pt}}l@{\extracolsep{\fill}}r@{\extracolsep{0pt}}c@{\extracolsep{0pt}}l@{\extracolsep{\fill}}r@{\extracolsep{0pt}}c@{\extracolsep{0pt}}l@{\extracolsep{\fill}}}
\hline\hline
 & \multicolumn{3}{c}{$W$+2jets} & \multicolumn{3}{c}{$W$+3jets} & \multicolumn{3}{c}{$W$+$\ge$4jets} \\
\hline
$Wb\bar{b}$ & 5.7&$\pm$&1.6 & 2.2&$\pm$&0.6 & 0.7&$\pm$&0.2 \\
Other bkg & 3.7&$\pm$&0.4 & 2.0&$\pm$&0.3 & 0.5&$\pm$&0.3 \\
\hline
Total bkg & 9.4&$\pm$&1.8 & 4.2&$\pm$&0.8 & 1.2&$\pm$&0.3 \\
$t\bar{t}\rightarrow \ell$+jets & 0.8&$\pm$&0.2 & 5.9&$\pm$&0.7 & 10.9&$\pm$&1.9 \\
\hline
Total expected & 10.2&$\pm$&1.9 & 10.1&$\pm$&1.2 & 12.1&$\pm$&2.0 \\
Observed & \multicolumn{3}{c}{15} & \multicolumn{3}{c}{9} & \multicolumn{3}{c}{14}\\
\hline\hline
\end{tabular*}
\end{table}

%% file: syst_table.tex
\begin{table}[b!]
 \caption{ \label{tab:x_emu}Summary of systematic uncertainties on $\sigma_{t\bar{t}}$.}
\begin{ruledtabular}
\begin{tabular}{lc}
    Source     &   {$\Delta\sigma_{t\bar{t}}$ (pb)}\\
\hline
$b$-tagging efficiency   & $+0.6 \; -0.5$ \\
Jet energy calibration   & $+0.5 \; -0.4$ \\
Background modeling      & $\pm 0.5$ \\
Lepton selections        & $+0.5 \; -0.4$ \\
Jet identification       & $+0.3 \; -0.2$ \\
Multijet background      & $+0.3 \; -0.2$ \\
Mis-tagging rate         & $\pm 0.1$ \\
\hline
Total                    & $+1.1 \; -1.0$ \\
\end{tabular}
\end{ruledtabular}
\end{table}

%% file: acknowledgement_paragraph_r2.tex
%
We thank the staffs at Fermilab and collaborating institutions, 
and acknowledge support from the 
DOE and NSF (USA),
CEA and CNRS/IN2P3 (France),
FASI, Rosatom and RFBR (Russia),
CAPES, CNPq, FAPERJ, FAPESP and FUNDUNESP (Brazil),
DAE and DST (India),
Colciencias (Colombia),
CONACyT (Mexico),
KRF (Korea),
CONICET and UBACyT (Argentina),
FOM (The Netherlands),
PPARC (United Kingdom),
MSMT (Czech Republic),
CRC Program, CFI, NSERC and WestGrid Project (Canada),
BMBF and DFG (Germany),
SFI (Ireland),
A.P.~Sloan Foundation,
Research Corporation,
Texas Advanced Research Program,
Alexander von Humboldt Foundation,
and the Marie Curie Fellowships.

%% file: paper_v.13.bbl
\begin{thebibliography}{99}
\bibitem{eta} Rapidity $y$ and pseudorapidity $\eta$ are defined as functions
of the polar angle $\theta$ and parameter $\beta$ as
$y(\theta,\beta) \equiv
{1 \over 2} \ln{[(1+\beta\cos{\theta})/(1-\beta\cos{\theta})]}$ and
$\eta(\theta) \equiv y(\theta,1)$, where $\beta$ is the ratio of a particle's 
momentum to its energy.
\bibitem{ip} Impact parameter is defined as the distance of closest 
approach ($d_{ca}$) of the track to the primary vertex in the 
plane transverse to the beamline. Impact parameter significance is defined as
$d_{ca}/\sigma_{d_{ca}}$, where $\sigma_{d_{ca}}$ is the uncertainty 
on $d_{ca}$.
\bibitem{dl} Decay length $L_{xy}$ is defined as the distance 
from the primary to the secondary vertex in the plane transverse to the beamline.
Decay length significance is defined as $L_{xy}/\sigma_{L_{xy}}$, where
$\sigma_{L_{xy}}$ is the uncertainty on $L_{xy}$.
\end{thebibliography}
